\documentclass[twocolumn,aps,prl,amsmath,amssymb]{revtex4-1}
\usepackage[latin9]{inputenc}
\usepackage{xcolor}
\usepackage{amsmath}
\usepackage{amssymb}
\usepackage{graphicx}
\usepackage{esint}

\makeatletter

\newcommand*\LyXThinSpace{\,\hspace{0pt}}
\DeclareFontEncoding{LGR}{}{}
\DeclareRobustCommand{\greektext}{%
  \fontencoding{LGR}\selectfont\def\encodingdefault{LGR}}
\DeclareRobustCommand{\textgreek}[1]{\leavevmode{\greektext #1}}
\ProvideTextCommand{\~}{LGR}[1]{\char126#1}

\usepackage{epsfig}
\usepackage{bm}
\usepackage{dcolumn}
\usepackage{color}

\makeatother

\begin{document}
\title{First-order Bose-Einstein condensation with three-body interacting
bosons}
\author{Hui Hu$^{1}$, Zeng-Qiang Yu$^{2,3}$, Jia Wang$^{1}$, and Xia-Ji
Liu$^{1}$}
\affiliation{$^{1}$Centre for Quantum Technology Theory, Swinburne University
of Technology, Melbourne, Victoria 3122, Australia}
\affiliation{$^{2}$Institute of Theoretical Physics, Shanxi University, Taiyuan
030006, China}
\affiliation{$^{3}$State Key Laboratory of Quantum Optics and Quantum Optics Devices,
Shanxi University, Taiyuan 030006, China}
\date{\today}
\begin{abstract}
Bose-Einstein condensation, observed in either strongly interacting
liquid helium or weakly interacting atomic Bose gases, is widely known
to be a second-order phase transition. Here, we predict a first-order
Bose-Einstein condensation in a cloud of harmonically trapped bosons
interacting with both attractive two-body interaction and repulsive
three-body interaction, characterized respectively by an $s$-wave
scattering length $a<0$ and a three-body scattering hypervolume $D>0$.
It happens when the harmonic trapping potential is weak, so with increasing
temperature the system changes from a low-temperature liquid-like
quantum droplet to a normal gas, and therefore experiences a first-order
liquid-to-gas transition. At large trapping potential, however, the
quantum droplet can first turn into a superfluid gas, rendering the
condensation transition occurred later from a superfluid gas to a
normal gas smooth. We determine a rich phase diagram and show the
existence of a tri-critical point, where the three phases - quantum
droplet, superfluid gas and normal gas - meet together. We argue that
an ensemble of spin-polarized tritium atoms could be a promising candidate
to observe the predicted first-order Bose-Einstein condensation, across
which the condensate fraction or central condensate density jumps
to zero and the surface-mode frequencies diverge.
\end{abstract}
\maketitle
Bose-Einstein condensation (BEC) is a ubiquitous quantum phenomenon
that occurs in a wide range of many-body systems, including liquid
helium \cite{Griffin1993}, atomic Bose gases \cite{Anderson1995,Dalfovo1999},
conventional and high-temperature superconductors \cite{Lee2006},
and even the hypothetical dark matter axions \cite{Sikivie2009}.
All the BEC transitions observed so far are of second-order and are
described by the model $F$ in the $XY$ universality class with $O(2)$
symmetry \cite{Hohenberg1977}. In strongly interacting liquid helium,
it is a continuous transition from a normal liquid (He-I) to a superfluid
liquid (He-II) across the $\lambda$-line \cite{Griffin1993}; while
in weakly interacting gaseous Bose systems of $^{87}$Rb atoms \cite{Anderson1995,Dalfovo1999},
it is a smooth transition from a normal gas to a superfluid gas. 

In this Letter, we propose that a first-order BEC transition from
a superfluid liquid to a normal gas could occur in weakly interacting
atomic Bose gases, when the two-body interaction is tuned to be attractive
(i.e., the $s$-wave scattering length $a<0$) and the resulting mean-field
collapse is arrested by a repulsive three-body interaction characterized
by the scattering hypervolume $D>0$. Our proposal is motivated by
the recent experimental realizations of liquid-like quantum droplets
\cite{Bottcher2021} in dipolar Bose-Einstein condensates \cite{FerrierBarbut2016,Schmitt2016,Chomaz2016}
or two-component Bose-Bose mixtures with attractive inter-species
interactions \cite{Cabrera2018,Semeghini2018,DErrico2019}, which
may experience a liquid-to-gas transition at nonzero temperature.
However, a careful examination of the temperature effect \cite{Wang2020a,Wang2020b}
indicates that the Lee-Huang-Yang (LHY) quantum fluctuation, which
is the key ingredient of the droplet formation \cite{Petrov2015,Minardi2019,Hu2020a,Hu2020b,Gu2020,Zin2021,Cui2021,Pan2021},
is too fragile to finite temperature. As a result, LHY droplets are
thermally destabilized far below the superfluid transition \cite{Wang2020a,Wang2020b}.
To overcome such a thermal instability, we resort to earlier cold-atom
proposals for quantum droplets based on the three-body repulsive interactions
\cite{Gammal2000,Bulgac2002,Blume2002,Gao2004,Beslic2009}, which
play the same role as LHY quantum fluctuations but are less sensitive
to temperature. These proposals regain considerable interest mostly
recently \cite{Zwerger2019,Mestrom2020}, owing to the brilliant idea
by Shina Tan and his co-worker that non-trivial three-body effects
can be expressed in terms of a single parameter of the hypervolume
$D$ \cite{Tan2008}, which can become positive and significant near
the zero-crossing of the two-body scattering length $a$ \cite{Zhu2017}.
Quantum droplets supported by the three-body interactions at zero
temperature have then been discussed \cite{Zwerger2019,Mestrom2020,Hu2020c}. 

Here, we address the finite-temperature properties of three-body interacting
bosons confined in three-dimensional (3D) harmonic traps, by using
the standard Hartree-Fock-Bogoliubov-Popov (HFB-Popov) theory \cite{Griffin1996,Hutchinson1997,Shi1998}.
At weak trapping potential, we find two phases, a low-temperature
quantum droplet and a high-temperature normal gas, separated by the
first-order BEC. While at large trapping potential, another superfluid
gas phase emerges and replaces the droplet phase. This leads to the
conventional smooth BEC transition between a superfluid gas and a
normal gas. An intriguing tri-critical point is formed in the phase
diagram, at which the quantum droplet, superfluid gas and normal gas
intersect. We explore the scenario of realizing such a tri-critical
point with ultracold atoms, for example, using a cloud of spin-polarized
tritium atoms \cite{Blume2002,Beslic2009}.

Incidentally, a similar tri-critical point has recently been discussed
by Dam Thanh Son and his collaborators for ultra quantum liquids formed
by a \emph{hypothetical} isotope of helium with nuclear mass less
than 4 atomic mass units \cite{Kora2020,Son2021}. Our results complement
their studies and take the unique advantage of the unprecedented controllability
and simplicity with ultracold atoms \cite{Dalfovo1999}. The liquid-to-gas
transition and BEC transition have also been considered in the context
of strongly-interacting matter of $\alpha$-particles \cite{Satarov2017}.
Therefore, it turns out that the first-order BEC transition predicted
in our work may find wide applications in diverse fields of physics,
ranging from atomic, molecular and optical physics, to condensed matter
physics, and to high-energy particle physics and nuclear physics.

\textit{Model Hamiltonian}. --- Three-body interacting bosons of
mass $M$ in 3D harmonic traps under consideration can be well described
by the model Hamiltonian, $\hat{H}=\int d\mathbf{r}\mathcal{H}(\mathbf{r})$,
with the Hamiltonian density,
\begin{equation}
\mathcal{H}=\hat{\psi}^{\dagger}\left[-\frac{\hbar^{2}\nabla^{2}}{2M}+V_{T}-\mu\right]\hat{\psi}+\frac{g}{2}\hat{\psi}^{\dagger2}\hat{\psi}^{2}+\frac{G}{6}\hat{\psi}^{\dagger3}\hat{\psi}^{3}.\label{eq:Hamiltonian}
\end{equation}
Here, $\hat{\psi}(\mathbf{r})$ and $\hat{\psi}^{\dagger}(\mathbf{r})$
are respectively annihilation and creation field operators of bosons,
$\mu$ is the chemical potential to be fixed by the total number of
atoms $N$, $g\equiv4\pi\hbar^{2}a/M<0$ and $G\equiv\hbar^{2}D/M>0$
are the attractive two-body and repulsive three-body interaction strengths,
respectively. The harmonic trapping potential $V_{T}(\mathbf{r})\equiv M\omega^{2}r^{2}/2$
is necessary, to prevent the atoms from escaping in the gas-like phase
or the finite-temperature self-evaporation in the droplet state \cite{Wang2020b,Petrov2015}.

\textit{HFB-Popov theory}. --- The model Hamiltonian at nonzero temperature
$T$ can be conveniently solved by the HFB-Popov theory \cite{Griffin1996,Hutchinson1997,Shi1998}.
We decompose $\hat{\psi}(\mathbf{r},t)\equiv\Phi(\mathbf{r})+\tilde{\psi}(\mathbf{r},t)$
into a condensate wave-function $\Phi(\mathbf{r})$ and a field operator
$\tilde{\psi}(\mathbf{r},t)$ for noncondensate atoms. From the equation
of motion for $\hat{\psi}(\mathbf{r},t)$, we deduce within the Popov
approximation \cite{SM}: (i) the generalized Gross-Pitaevskii equation
(GPE) for the condensate wave-function, 
\begin{equation}
\mathcal{\hat{L}}\Phi(\mathbf{r})=\mu\Phi(\mathbf{r}),\label{eq:GPE}
\end{equation}
where we have defined the operator,
\[
\mathcal{\hat{L}}\equiv-\frac{\hbar^{2}\nabla^{2}}{2M}+V_{T}+g\left(n_{c}+2\tilde{n}\right)+G\left(\frac{n_{c}^{2}}{2}+3n_{c}\tilde{n}+3\tilde{n}^{2}\right),
\]
and $n_{c}(\mathbf{r})\equiv\left|\Phi(\mathbf{r})\right|^{2}$ and
$\tilde{n}(\mathbf{r})\equiv\bigl\langle\tilde{\psi}^{\dagger}(\mathbf{r})\tilde{\psi}(\mathbf{r})\bigr\rangle$
are the condensate and noncondensate densities, respectively; and
(ii) the coupled HFB-Popov equations for the $\eta$-th quasi-particle
wave-functions $u_{\eta}$ and $v_{\eta}$ with energy $E_{\eta}>0$,
\begin{equation}
\left[\begin{array}{cc}
\hat{\mathcal{L}}-\mu+\mathcal{\hat{M}} & \mathcal{\hat{M}}\\
\mathcal{\hat{M}} & \hat{\mathcal{L}}-\mu+\mathcal{\hat{M}}
\end{array}\right]\left[\begin{array}{c}
u_{\eta}\left(\mathbf{r}\right)\\
v_{\eta}\left(\mathbf{r}\right)
\end{array}\right]=E_{\eta}\left[\begin{array}{c}
+u_{\eta}\left(\mathbf{r}\right)\\
-v_{\eta}\left(\mathbf{r}\right)
\end{array}\right],\label{eq:HFBP}
\end{equation}
where the operator $\hat{\mathcal{M}}\equiv gn_{c}+G(n_{c}^{2}+3n_{c}\tilde{n}).$
Once the quasi-particle wave-functions are obtained, the noncondensate
density can be calculated according to,
\begin{equation}
\tilde{n}\left(\mathbf{r}\right)=\tilde{n}_{\textrm{qd}}\left(\mathbf{r}\right)+\sum_{\eta}\frac{\left|u_{\eta}\left(\mathbf{r}\right)\right|^{2}+\left|v_{\eta}\mathbf{\left(r\right)}\right|^{2}}{e^{\beta E_{\eta}}-1},
\end{equation}
where $\tilde{n}_{\textrm{qd}}(\mathbf{r})=\sum_{\eta}\left|v_{\eta}\mathbf{(r})\right|^{2}$
is the depletion to the condensate arising from quantum fluctuations
and $\beta\equiv1/(k_{B}T)$. In the absence of the three-body interaction,
i.e., $G=0$, Eq. (\ref{eq:GPE}) and Eq. (\ref{eq:HFBP}) recover
the well known HFB-Popov theory of a weakly interacting Bose gas \cite{Hutchinson1997}.
In the normal state ($n_{c}=0$ and the total density $n=\tilde{n}$),
Eq. (3) instead describes the single-particle motion under a mean-field
Hartree-Fock interaction potential $2gn+3Gn^{2}$. 

The Popov approximation amounts to neglecting the anomalous correlation
$\tilde{m}(\mathbf{r})\equiv\bigl\langle\tilde{\psi}(\mathbf{r})\tilde{\psi}(\mathbf{r})\bigr\rangle$,
which is a higher-order effect beyond mean-field \cite{Griffin1996,Shi1998}.
It ensures the \emph{gapless} phonon spectrum in the homogeneous limit
\cite{Griffin1996}, where the chemical potential is given by $\mu=g(n_{c}+2\tilde{n})+G(n_{c}^{2}/2+3n_{c}\tilde{n}+3\tilde{n}^{2})$
and hence $\mathcal{\hat{L}}_{k=0}=\mu$. However, it is worth noting
that, at low temperature the anomalous correlation is at the same
order as the quantum depletion in magnitude. For consistency, therefore,
we neglect the quantum depletion in the noncondensate density. This
treatment is reasonable, since the quantum depletion is typical about
ten percent \cite{SM} and its absence recovers the standard GPE for
condensate wave-function that has been widely adopted in the previous
studies \cite{Gammal2000,Bulgac2002,Mestrom2020,Hu2020c}. We also
note that, the Popov approximation leads to an \emph{artificial} first-order
superfluid transition with a few percent jump in the condensate density
\cite{Shi1998}. This drawback has nothing to do with the first-order
BEC transition predicted in our work, where the sudden jump in the
central condensate density at the transition is almost $100\%$ \cite{SM}.

\textit{Numerical calculations}. --- We have iteratively solved Eq.
(\ref{eq:GPE}) and Eq. (\ref{eq:HFBP}) in a self-consistent way,
with the chemical potential $\mu$ determined by the number equation
$N=\int d\mathbf{r}[n_{c}(\mathbf{r})+\tilde{n}(\mathbf{r})]=N_{c}+N_{\textrm{th}}$.
To ease the numerical workload, it is useful to introduce the re-scaled
units for length $\mathbf{\bar{r}}\equiv\mathbf{r}/\xi$, density
$\bar{n}\equiv n/n_{0}$, and energy $\bar{E}_{\eta}\equiv M\xi^{2}E_{\eta}/\hbar^{2}$,
where $\xi=\sqrt{D/(6\pi^{2}a^{2})}$ is the length scale and $n_{0}=6\pi\left|a\right|/D$
is the equilibrium density of zero-temperature quantum droplets \cite{SM},
so the two interaction strengths now become \emph{dimensionless}:
$gn_{0}M\xi^{2}/\hbar^{2}\equiv\bar{g}=-4$ and $Gn_{0}^{2}M\xi^{2}/\hbar^{2}\equiv\bar{G}=+6$.
The reduced number of particles is given by $\bar{N}=N/(n_{0}\xi^{3})$
with $n_{0}\xi^{3}=\sqrt{D/6}/(\pi^{2}a^{2})$. Hereafter, without
any confusion we shall remove the bar in the re-scale units. 

At zero temperature, the use of the re-scaled units leads to a simple
GPE equation, $[-\nabla^{2}/2+\omega^{2}r^{2}/2-4\Phi^{2}+3\Phi^{4}]\Phi=\mu\Phi$,
which has been well understood \cite{Gammal2000,Hu2020c}. For instance,
without harmonic traps ($\omega=0$), the GPE allows a self-bound
quantum droplet with $\Phi\simeq1$ and $\mu\simeq-1$ for large reduced
number of particles $N\gg1$ \cite{Bulgac2002,Hu2020c}. The droplet
state is robust below a characteristic trapping frequency, i.e., $\omega\lesssim\sqrt{2}(4\pi/3N)^{1/3}\simeq2.3N^{-1/3}$
\cite{Hu2020c}. It is easy to see that, if we neglect the finite-size
effect for large number of particles, the properties of the system
depend on the product $N^{1/3}\omega$, as in a weakly interacting
Bose gas \cite{Dalfovo1999}. Most numerical calculations in this
work are therefore carried out for $N=1000$. We also vary $N$ in
the range $[125,8000]$ and find no sizable finite-size effect.

In the case of quantum droplets the finite-temperature HFB-Popov equations
are generally more challenging to solve than that of a weakly interacting
Bose gas \cite{Hutchinson1997}. For the technical aspects of our
numerical calculations, we refer to Supplemental Material for details
\cite{SM}.

\begin{figure}[t]
\begin{centering}
\includegraphics[width=0.45\textwidth]{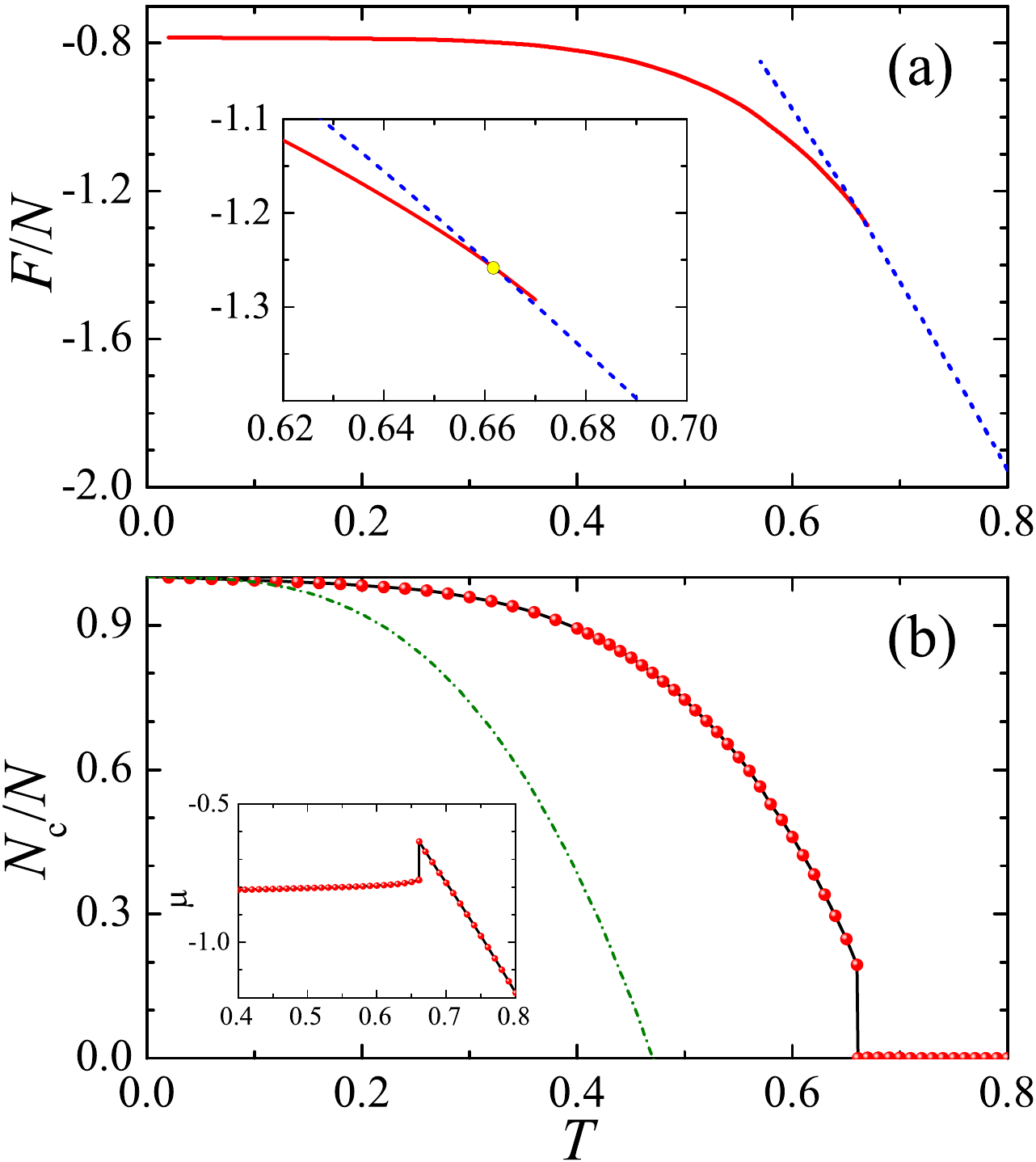}
\par\end{centering}
\caption{\label{fig1_1stBEC} First-order BEC transition at $N^{1/3}\omega=0.5$.
(a) Temperature dependence of the free energy for the condensed-state
solution (red solid line) and for the normal-state solution (blue
dashed line). The inset highlights the transition at $T_{c}\simeq0.661$.
(b) The condensate fraction as a function of temperature. The green
dot-dashed line shows $N_{c}/N=1-(T/T_{c0})^{3}$, where $T_{c0}\simeq0.9405N^{1/3}\omega$
is the transition temperature of an ideal Bose gas in harmonic traps
\cite{Dalfovo1999}. The inset shows a jump in the chemical potential
at $T_{c}$.}
\end{figure}

\textit{First-order BEC transition}. --- Let us first consider the
finite-temperature thermodynamics at small effective trapping frequency,
i.e., $N^{1/3}\omega=0.5$, as reported in Fig. \ref{fig1_1stBEC}.
Remarkably, near the superfluid transition we always find two possible
solutions: one comes with a significant condensate fraction, while
the other is completely a normal state. To identify which one is the
ground-state solution, we calculate the free energy $F=\Omega+\mu N$,
where the thermodynamic potential $\Omega$ takes the form \cite{SM},
\begin{eqnarray}
\Omega & = & \sum_{\eta}\frac{\ln\left[1-e^{-\beta E_{\eta}}\right]}{\beta}+\int d\mathbf{r}\Phi^{*}\left[-\frac{\hbar^{2}\nabla^{2}}{2M}+V_{T}-\mu\right]\Phi\nonumber \\
 &  & +\int d\mathbf{r}\left[\frac{gn_{c}^{2}}{2}-g\tilde{n}^{2}+G\left(\frac{n_{c}^{3}}{6}-3n_{c}\tilde{n}^{2}-2\tilde{n}^{3}\right)\right].
\end{eqnarray}
It is readily seen from Fig. \ref{fig1_1stBEC}(a), the free energies
of the two solutions intersect at $T_{c}\simeq0.661$ with different
slope, clearly indicating a first-order BEC transition. Consequently,
the condensate fraction suddenly drops to zero at $T_{c}$, as shown
in Fig. \ref{fig1_1stBEC}(b). It is also significantly larger than
the ideal gas result for non-interacting bosons in harmonic traps,
i.e., $N_{c}/N=1-(T/T_{c0})^{3}$, where $T_{c0}=\omega[N/\zeta(3)]^{1/3}$
with the Zeta function $\zeta(3)\simeq1.202$ \cite{Dalfovo1999}.
We find that with increasing temperature the non-condensate fraction
increases exponentially slowly, compared with the usual power-law
$T^{3}$ behavior in the gas-like phase. This slow increase is due
to the discrete excitation spectrum of the self-bound quantum droplet,
which persists even in the absence of the trapping potential \cite{Hu2020c}.
We note that, the sudden disappearance of the condensate fraction
is correlated with a jump in the chemical potential, as plotted in
the inset of Fig. \ref{fig1_1stBEC}(b). The observation of a first-order
BEC transition at small trapping frequency is the main result of our
work.

\begin{figure}[t]
\begin{centering}
\includegraphics[width=0.45\textwidth]{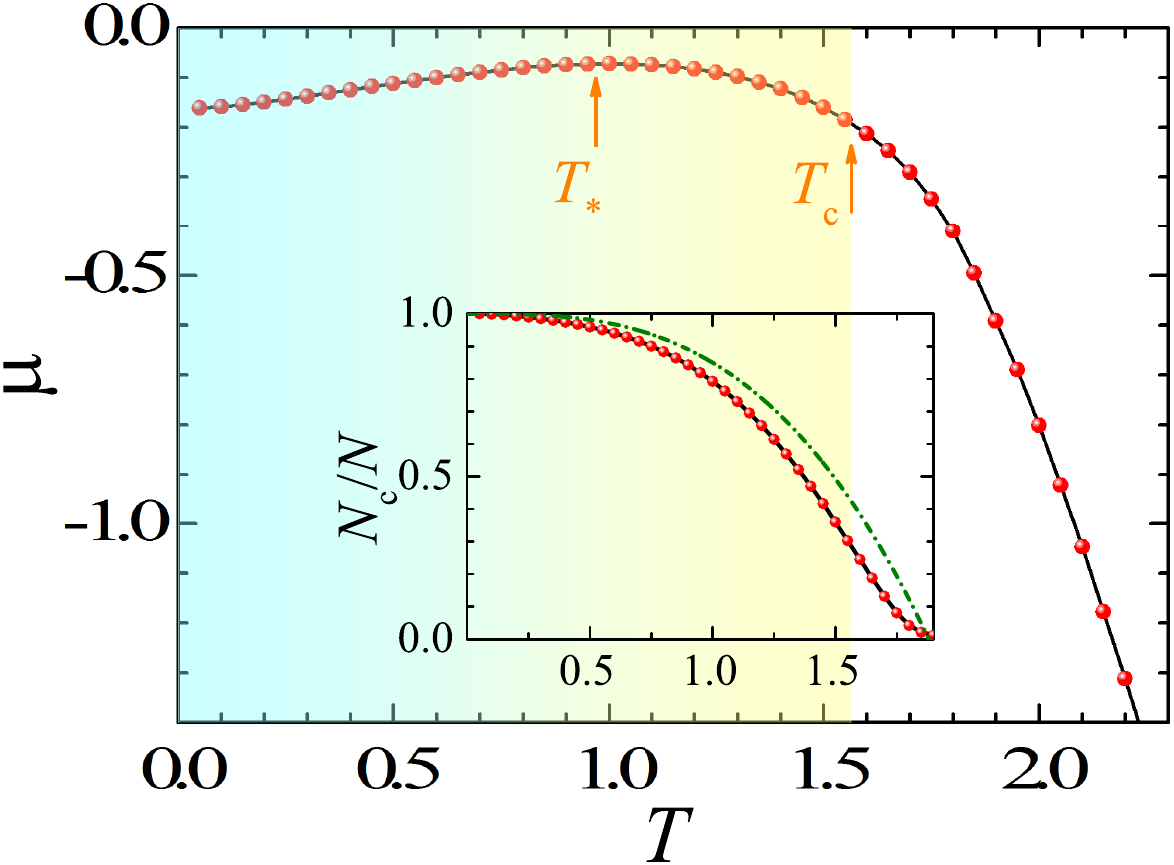}
\par\end{centering}
\caption{\label{fig2_2ndBEC} Second-order BEC transition at $N^{1/3}\omega=2.0$.
The chemical potential is plotted as a function of temperature. It
is a smooth function and the BEC transition occurs at $T_{c}\simeq1.563$.
The inset shows the condensate fraction and the green dot-dashed line
is the ideal gas prediction, $N_{c}/N=1-(T/T_{c0})^{3}$.}
\end{figure}

At large trapping frequency, the situation dramatically changes. A
typical case of $N^{1/3}\omega=2.0$ is presented in Fig. \ref{fig2_2ndBEC}.
Both chemical potential and condensate fraction change smoothly when
temperature increases, suggesting a second-order superfluid phase
transition. We interpret it as a transition from a superfluid gas
to a normal gas, and therefore use the standard approach to determine
a critical temperature $T_{c}\simeq1.563$, at which the condensate
fraction should change most significantly (i.e., $d^{2}N_{c}/dT^{2}=0$)
\cite{Hu2003}. Our interpretation follows the two observations that
the chemical potential is a decreasing function of temperature near
the superfluid transition and the condensate fraction lies systematically
below the ideal gas prediction (i.e., the green dot-dashed line),
both of which are the key features of the second-order transition
of a weakly interacting Bose gas \cite{Dalfovo1999}. Interestingly,
at low temperature the chemical potential is rather an increasing
function of temperature up to a turning point (indicated by $T_{*}\simeq1.0$
in the figure), which is consistent with the picture of a quantum
droplet \cite{muQuantumDroplet}. Thus, the system seems to cross
from a liquid-like droplet over to a gas-like phase at the characteristic
temperature $T_{*}$.

\begin{figure}[t]
\begin{centering}
\includegraphics[width=0.45\textwidth]{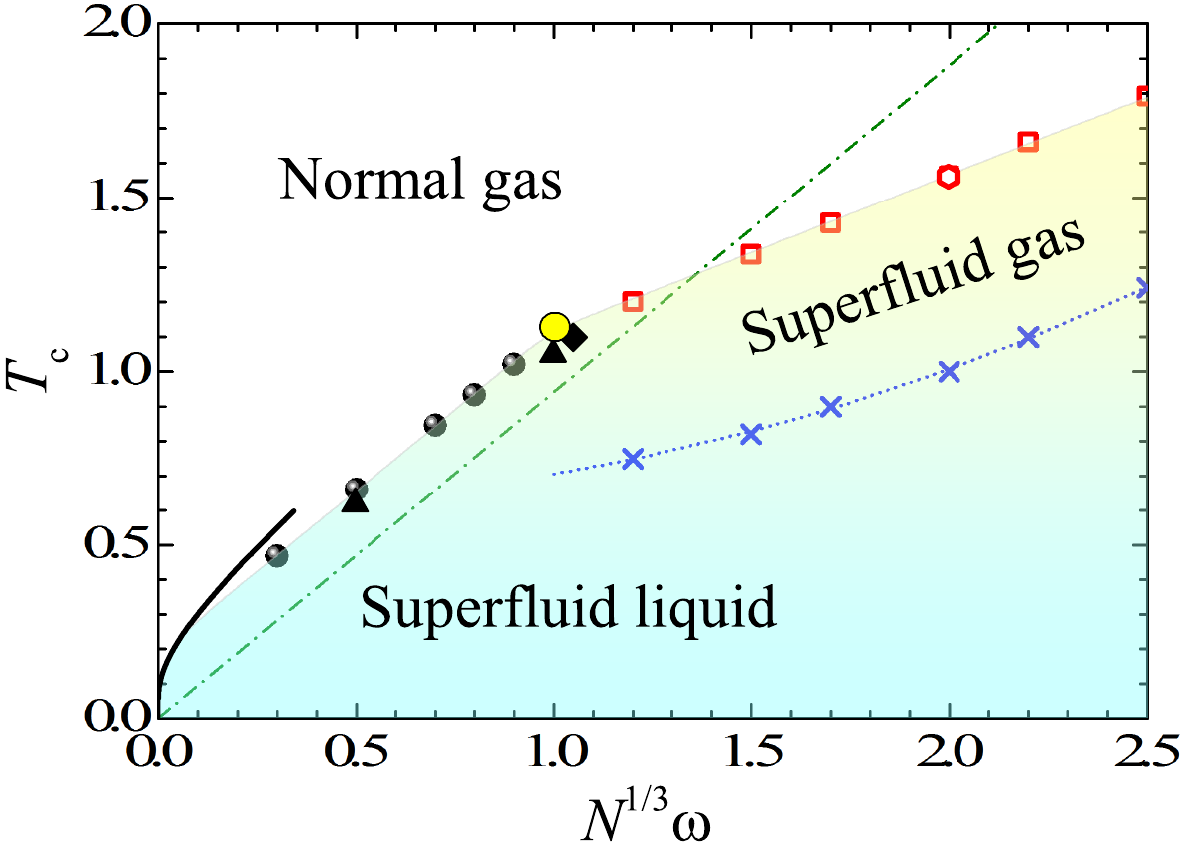}
\par\end{centering}
\caption{\label{fig3_phasediagram} Critical temperature $T_{c}$ (closed and
open symbols) and crossover temperature $T_{*}$ (crosses) as a function
of $N^{1/3}\omega$. The thick solid line is the critical temperature
predicted at small trapping frequency, i.e., $N^{1/3}\omega=T_{c}\exp\left[-1/(3T_{c})\right]$.
The green dot-dashed line shows $T_{c0}\simeq0.9405N^{1/3}\omega$.
The dotted line that connects the crossover temperature $T_{*}$ is
the guide for the eye. A tri-critical point is highlighted by the
yellow dot. Typically, we take $N=1000$. The change in the reduced
number of particles (i.e., the triangles for $N=125$ and diamonds
for $N=3375$ and hexagons for $N=8000$) does not lead to sizable
difference to $T_{c}$.}
\end{figure}

\textit{Phase diagram}. --- By calculating the critical temperature
$T_{c}$ and crossover temperature $T_{*}$ at different effective
trapping frequency $N^{1/3}\omega$, we determine a phase diagram
in Fig. \ref{fig3_phasediagram}. An intriguing tri-critical point
appears at $(N^{1/3}\omega)_{\textrm{tri}}\simeq1.0$ and $T_{\textrm{tri}}\simeq1.13$,
where the droplet phase (i.e., superfluid liquid), superfluid gas
and normal gas intersect with others. Below the tri-critical trapping
frequency, i.e., $N^{1/3}\omega<1.0$, a superfluid liquid turns into
a normal gas via a first-order transition (black solid symbols); while
at $N^{1/3}\omega>1.0$, with increasing temperature the superfluid
liquid first becomes a superfluid gas at the crossover temperature
$T_{*}$ (crosses) and then turns into a normal gas via a smooth second-order
phase transition (red empty symbols). Note that, with decreasing $N^{1/3}\omega$
the crossover temperature $T_{*}$ does not converge to the tri-critical
point. This is probably due to the difficulty of defining an appropriate
crossover temperature in a finite-size system close to the tri-critical
point \cite{SM}.

At vanishingly small trapping frequency (i.e., $N^{1/3}\omega\rightarrow0$),
the critical temperature can be analytically derived \cite{SM}. We
find that the relation $N^{1/3}\omega=T_{c}\exp\left[-1/(3T_{c})\right]$
and hence $T_{c}\sim-\ln^{-1}(N\omega^{3})$. Due to the logarithmic
dependence, $T_{c}$ could remain sizable at \emph{negligible} trapping
frequency. As the self-evaporation rate of the droplet is very slow
at low temperature \cite{Barranco2006}, it seems likely to find a
\emph{self-bound} quantum droplet at small but nonzero critical temperature
(i.e., $T_{c}\simeq0.1\left|\mu\right|$, where the chemical potential
$\mu\simeq-1$ sets the energy scale), when we gradually remove the
external harmonic trapping potential.

\begin{figure}[t]
\begin{centering}
\includegraphics[width=0.45\textwidth]{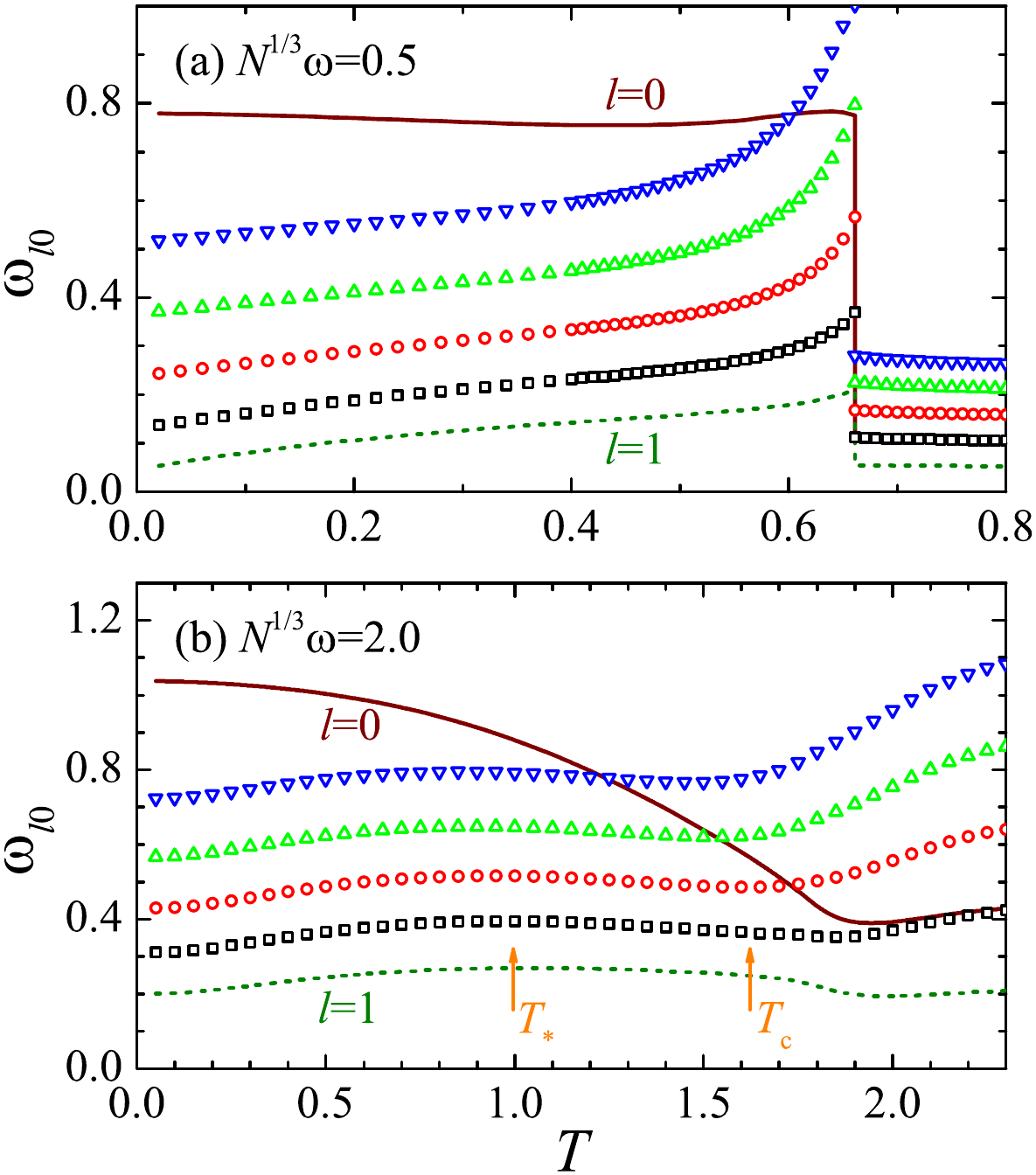}
\par\end{centering}
\caption{\label{fig4_cm} The mode frequencies $\omega_{l,n=0}$ of the breathing
mode ($l=0$, solid lines), dipole mode ($l=1$, dotted lines) and
surface modes ($l=2,3,4,5$, symbols from bottom to top) as a function
of temperature $T$ at $N^{1/3}\omega=0.5$ (a) and $N^{1/3}\omega=2.0$
(b). In (b), the two arrows indicate the crossover temperature $T_{*}$
and the superfluid transition temperature $T_{c}$, respectively.
We note that, the dipole mode frequency $\omega_{10}$ is not precisely
the trapping frequency $\omega$, due to the mean-field interaction
potential in the HFB-Popov theory \cite{Hutchinson1997}.}
\end{figure}

\textit{Observation of the first-order transition}. --- The predicted
first-order BEC transition can be straightforwardly probed from the
jump in the condensate fraction, or more readily from the discontinuity
in the central condensate density, as we discuss in detail in Supplemental
Material \cite{SM}. For small effective trapping frequency, we find
that the central condensate density is nearly unchanged below the
superfluid transition and suddenly drops to zero right at the critical
temperature $T_{c}$ \cite{SM}.

Alternatively, we may probe the first-order transition by measuring
the collective excitations of the system. For small trapping frequency,
the quantum droplet features peculiar surface modes known as ripplons
\cite{Petrov2015,Hu2020c,Baillie2017}. We find that the ripplon mode
frequency $\omega_{l\geq2,n=0}$ diverges towards the first-order
BEC transition at $T_{c}$, as shown in Fig. \ref{fig4_cm}(a). In
contrast, at large trapping frequency there seems to be local minimum
in the ripplon mode frequencies near the second-order BEC transition,
as can be seen from Fig. \ref{fig4_cm}(b). The ripplon mode frequencies
also show a characteristic local maximum at about the crossover temperature
$T_{*}$, which might experimentally be used to locate the smooth
crossover from a quantum droplet to a superfluid gas.

\textit{Spin-polarized tritium atoms}. ---In previous theoretical
studies of the quantum droplet stabilized by three-body interactions,
spin-polarized tritium condensate has been suggested to be a good
candidate \cite{Blume2002,Beslic2009}. We have further explored this
possibility by investigating the hypervolume $D$ of tritium atoms,
inspired by the recent universality work by Mestrom \textsl{et al.}
\cite{Mestrom2020}. In the absence of Feshbach resonances, the $s$-wave
scattering length $a\simeq-11.92r_{\textrm{vdW}}$ and the trimer
energy $E_{T}\simeq-3.50\times10^{-3}E_{\textrm{vdW}}$, where $r_{\textrm{vdW}}$
and $E_{\textrm{vdW}}$ are the characteristic length and energy of
the van-der-Waals potential of tritium atoms \cite{Blume2002}. The
trimer energy differs slightly from the universal value $E_{T,\textrm{uni}}\simeq-4.45\times10^{-3}E_{\textrm{vdW}}$
\cite{WangPhDThesis}, suggesting that the universal formalism for
the hypervolume $D$ given by Mestrom \textsl{et al.} \cite{Mestrom2020}
could be used. As there is no decay dimer channel, so $D$ is purely
real, without the annoying three-body loss \cite{Zwerger2019}. We
can further apply Feshbach resonances to tune both $a$ and $D$ to
realize a cloud of three-body weakly interacting tritium atoms. Cooling
and trapping spin-polarized hydrogen and deuterium have recently been
realized using Zeeman decelerator \cite{Hogan2008,Wiederkehr2010}.
The same technique can be applied to tritium \cite{ClarkBSThesis}.

\textit{Conclusions}. --- We have proposed that three-body interacting
bosons may experience a first-order Bose-Einstein condensation in
a weak harmonic trapping potential, in sharp contrast to the conventional
smooth condensation transition observed so far. This first-order transition
can be unambiguously probed from the sudden jump in the central density
and the divergent ripplon mode frequency at the critical temperature.
We have suggested that spin-polarized tritium atoms could be a promising
candidate for observing the predicted first-order Bose-Einstein condensation
transition.
\begin{acknowledgments}
This research was supported by the Australian Research Council's (ARC)
Discovery Program, Grant No. DP170104008 (H.H.), Grants No. DE180100592
and No. DP190100815 (J.W.), and Grant No. DP180102018 (X.-J.L), and
also by the National Natural Science Foundation of China (NSFC) under
Grant No. 11674202 (Z.Q.Y.) and the fund for Shanxi ``1331 KSC Project\textquoteright \textquoteright{}
(Z.Q.Y.).
\end{acknowledgments}

\appendix

\section{The HFB-Popov theory}

We start with the model Hamiltonian for $N$ three-body interacting
bosons of mass $M$ in 3D harmonic traps $V_{T}(\mathbf{r})\equiv M\omega^{2}r^{2}/2$,
\begin{equation}
\hat{H}=\int d\mathbf{r}\mathcal{H}(\mathbf{r}),
\end{equation}
where the Hamiltonian density is,
\begin{equation}
\mathcal{H}=\hat{\psi}^{\dagger}\left[-\frac{\hbar^{2}\nabla^{2}}{2M}+V_{T}-\mu\right]\hat{\psi}+\frac{g}{2}\hat{\psi}^{\dagger2}\hat{\psi}^{2}+\frac{G}{6}\hat{\psi}^{\dagger3}\hat{\psi}^{3}.\label{eq:Hamiltonian-1}
\end{equation}
Here, $\hat{\psi}(\mathbf{r})$ and $\hat{\psi}^{\dagger}(\mathbf{r})$
are respectively the annihilation and creation field operators of
bosons, $\mu$ is the chemical potential, and the attractive two-body
and repulsive three-body interaction strengths are given by,
\begin{eqnarray}
g & \equiv & \frac{4\pi\hbar^{2}a}{M}<0,\\
G & \equiv & \frac{\hbar^{2}D}{M}>0,
\end{eqnarray}
respectively. To derive the HFB-Popov equations, we follow the procedure
in the seminal work by Allan Griffin \cite{Griffin1996} and decompose
the field operator $\hat{\psi}(\mathbf{r})$ into a condensate wave-function
$\Phi(\mathbf{r})$ and a field operator $\tilde{\psi}(\mathbf{r})$
for noncondensate atoms, i.e., $\hat{\psi}(\mathbf{r},t)\equiv\Phi(\mathbf{r})+\tilde{\psi}(\mathbf{r},t)$
and its Hermitian form $\hat{\psi}^{\dagger}(\mathbf{r},t)\equiv\Phi^{*}(\mathbf{r})+\tilde{\psi}^{\dagger}(\mathbf{r},t)$.

In the exact Heisenberg equation of motion for $\hat{\psi}(\mathbf{r})$,\begin{widetext}
\begin{equation}
i\hbar\frac{\partial\hat{\psi}\left(\mathbf{r},t\right)}{\partial t}=\left[-\frac{\hbar^{2}\nabla^{2}}{2M}+V_{T}-\mu\right]\hat{\psi}\left(\mathbf{r},t\right)+g\hat{\psi}^{\dagger}\hat{\psi}\hat{\psi}\left(\mathbf{r},t\right)+\frac{G}{2}\hat{\psi}^{\dagger}\hat{\psi}^{\dagger}\hat{\psi}\hat{\psi}\hat{\psi}\left(\mathbf{r},t\right),\label{eq:totEoM}
\end{equation}
let us take the thermal average and obtain, 
\begin{equation}
\left[-\frac{\hbar^{2}\nabla^{2}}{2M}+V_{T}-\mu\right]\Phi(\mathbf{r})+g\bigl\langle\hat{\psi}^{\dagger}\hat{\psi}\hat{\psi}\bigr\rangle+\frac{G}{2}\bigl\langle\hat{\psi}^{\dagger}\hat{\psi}^{\dagger}\hat{\psi}\hat{\psi}\hat{\psi}\bigr\rangle=0.\label{eq:BECEoM}
\end{equation}
The last two terms in the above equation of motion can be treated
in the \emph{self-consistent} mean-field approximation \cite{Griffin1996},
namely ($\tilde{n}\equiv\bigl\langle\tilde{\psi}^{\dagger}\tilde{\psi}\bigr\rangle$
and $\tilde{m}\equiv\bigl\langle\tilde{\psi}\tilde{\psi}\bigr\rangle$),
\begin{equation}
\bigl\langle\hat{\psi}^{\dagger}\hat{\psi}\hat{\psi}\bigr\rangle=\Phi^{2}\Phi^{*}+\Phi^{*}\bigl\langle\tilde{\psi}\tilde{\psi}\bigr\rangle+2\Phi\bigl\langle\tilde{\psi}^{\dagger}\tilde{\psi}\bigr\rangle=\Phi^{2}\Phi^{*}+\Phi^{*}\tilde{m}+2\Phi\tilde{n}
\end{equation}
and
\begin{eqnarray}
\bigl\langle\hat{\psi}^{\dagger}\hat{\psi}^{\dagger}\hat{\psi}\hat{\psi}\hat{\psi}\bigr\rangle & = & \Phi^{3}\Phi^{*2}+3\Phi\Phi^{*2}\bigl\langle\tilde{\psi}\tilde{\psi}\bigr\rangle+6\Phi^{2}\Phi^{*}\bigl\langle\tilde{\psi}^{\dagger}\tilde{\psi}\bigr\rangle+\Phi^{3}\bigl\langle\tilde{\psi}^{\dagger}\tilde{\psi}^{\dagger}\bigr\rangle+2\Phi^{*}\bigl\langle\tilde{\psi}^{\dagger}\tilde{\psi}\tilde{\psi}\tilde{\psi}\bigr\rangle+3\Phi\bigl\langle\tilde{\psi}^{\dagger}\tilde{\psi}^{\dagger}\tilde{\psi}\tilde{\psi}\bigr\rangle,\nonumber \\
 & = & \Phi^{3}\Phi^{*2}+3\Phi\Phi^{*2}\tilde{m}+6\Phi^{2}\Phi^{*}\tilde{n}+\Phi^{3}\tilde{m}^{*}+6\Phi^{*}\tilde{n}\tilde{m}+3\Phi\left(2\tilde{n}^{2}+\tilde{m}\tilde{m}^{*}\right).
\end{eqnarray}
Here, we have decoupled the four-field-operator terms according to
the Wick theorem,
\begin{eqnarray}
\bigl\langle\tilde{\psi}^{\dagger}\tilde{\psi}\tilde{\psi}\tilde{\psi}\bigr\rangle & = & 3\tilde{n}\tilde{m},\\
\bigl\langle\tilde{\psi}^{\dagger}\tilde{\psi}^{\dagger}\tilde{\psi}\tilde{\psi}\bigr\rangle & = & 2\tilde{n}^{2}+\tilde{m}\tilde{m}^{*}.
\end{eqnarray}
Therefore, we find the time-independent generalized Gross-Pitaevskii
(GP) equation ($n_{c}\equiv\left|\Phi(\mathbf{r})\right|^{2}$),
\begin{equation}
{\color{black}\left[-\frac{\hbar^{2}\nabla^{2}}{2M}+V_{T}-\mu\right]\Phi+g\left(n_{c}+2\tilde{n}\right)\Phi+{\normalcolor {\color{red}g\tilde{m}\Phi^{*}}}+\frac{G}{2}\left(n_{c}^{2}+6n_{c}\tilde{n}+6\tilde{n}^{2}+{\color{red}\Phi^{2}\tilde{m}^{*}+3\tilde{m}\tilde{m}^{*}}\right)\Phi+{\color{red}\frac{3G}{2}\left(n_{c}\tilde{m}+2\tilde{n}\tilde{m}\right)\Phi^{*}}=0.}
\end{equation}
It is well-known that a nonzero anomalous correlation $\tilde{m}\equiv\bigl\langle\tilde{\psi}\tilde{\psi}\bigr\rangle$
gives rise to a \emph{gapped} excitation spectrum \cite{Griffin1996,Shi1998},
even in the absence of the three-body interacting term, which is unphysical.
Therefore, we take the Popov approximation $\tilde{m}=0$ and $\tilde{m}^{*}=0$
everywhere in the generalized GP equation. This leads to the form,
\begin{equation}
\mathcal{\hat{L}}\Phi\left(\mathbf{r}\right)\equiv\left[-\frac{\hbar^{2}\nabla^{2}}{2M}+V_{T}+g\left(n_{c}+2\tilde{n}\right)+G\left(\frac{n_{c}^{2}}{2}+3n_{c}\tilde{n}+3\tilde{n}^{2}\right)\right]\Phi\left(\mathbf{r}\right)=\mu\Phi\mathbf{\left(r\right)}.\label{eq:GPE-1}
\end{equation}

The generalized Hartree-Fock-Bogoliubov equation for quasi-particles
may be derived from the equation of motion for the field operator
$\tilde{\psi}(\mathbf{r})$ of noncondensate atoms \cite{Griffin1996},
which can be obtained by subtracting Eq. (\ref{eq:BECEoM}) from Eq.
(\ref{eq:totEoM}). It can also be equivalently derived by expanding
the model Hamiltonian to the quadratic terms of $\tilde{\psi}^{\dagger}$
and $\tilde{\psi}$ \cite{Griffin1996}. This alternative derivation
is useful to determine the expressions of the thermodynamic potential
and free energy. Therefore, let us describe it in detail. We first
consider the two-body interaction term,

\begin{eqnarray}
\hat{\psi}^{\dagger2}\hat{\psi}^{2} & = & \Phi^{2}\Phi^{*2}+\Phi^{2}\tilde{\psi}^{\dagger}\tilde{\psi}^{\dagger}+4\Phi\Phi^{*}\tilde{\psi}^{\dagger}\tilde{\psi}+\Phi^{*2}\tilde{\psi}\tilde{\psi}+\tilde{\psi}^{\dagger}\tilde{\psi}^{\dagger}\tilde{\psi}\tilde{\psi},\nonumber \\
 & = & \left[\left(\Phi^{2}+\tilde{m}\right)\tilde{\psi}^{\dagger}\tilde{\psi}^{\dagger}+\textrm{h.c.}\right]+4\left(n_{c}+\tilde{n}\right)\tilde{\psi}^{\dagger}\tilde{\psi}+n_{c}^{2}-\left(2\tilde{n}^{2}+\tilde{m}\tilde{m}^{*}\right).
\end{eqnarray}
In the second line of the above equation, we have taken the self-consistent
mean-field approximation to $\tilde{\psi}^{\dagger}\tilde{\psi}^{\dagger}\tilde{\psi}\tilde{\psi}$,
i.e.,
\begin{equation}
\tilde{\psi}^{\dagger}\tilde{\psi}^{\dagger}\tilde{\psi}\tilde{\psi}=\tilde{m}\tilde{\psi}^{\dagger}\tilde{\psi}^{\dagger}+4\tilde{n}\tilde{\psi}^{\dagger}\tilde{\psi}+\tilde{m}^{*}\tilde{\psi}\tilde{\psi}-\left(2\tilde{n}^{2}+\tilde{m}\tilde{m}^{*}\right).
\end{equation}
The three-body interaction term can be treated in a similar way. We
find that,
\begin{equation}
\hat{\psi}^{\dagger3}\hat{\psi}^{3}=n_{c}^{3}+\left(3\Phi^{3}\Phi^{*}\tilde{\psi}^{\dagger}\tilde{\psi}^{\dagger}+\textrm{h.c.}\right)+9n_{c}^{2}\tilde{\psi}^{\dagger}\tilde{\psi}+9n_{c}\tilde{\psi}^{\dagger}\tilde{\psi}^{\dagger}\tilde{\psi}\tilde{\psi}+\left(3\Phi^{2}\tilde{\psi}^{\dagger}\tilde{\psi}^{\dagger}\tilde{\psi}^{\dagger}\tilde{\psi}+\textrm{h.c.}\right)+\tilde{\psi}^{\dagger3}\tilde{\psi}^{3}.
\end{equation}
By inserting the mean-field decoupling, i.e., 
\begin{eqnarray}
\tilde{\psi}^{\dagger}\tilde{\psi}^{\dagger}\tilde{\psi}^{\dagger}\tilde{\psi} & = & 3\tilde{n}\tilde{\psi}^{\dagger}\tilde{\psi}^{\dagger}+3\tilde{m}^{*}\tilde{\psi}^{\dagger}\tilde{\psi}-3\tilde{n}\tilde{m}^{*},\\
\tilde{\psi}^{\dagger3}\tilde{\psi}^{3} & = & 9\left[\tilde{n}\tilde{m}\tilde{\psi}^{\dagger}\tilde{\psi}^{\dagger}+\left(2\tilde{n}^{2}+\tilde{m}\tilde{m}^{*}\right)\tilde{\psi}^{\dagger}\tilde{\psi}+\tilde{n}\tilde{m}^{*}\tilde{\psi}\tilde{\psi}\right]-\left(12\tilde{n}^{3}+18\tilde{n}\tilde{m}\tilde{m}^{*}\right),
\end{eqnarray}
we obtain that,
\begin{eqnarray}
\hat{\psi}^{\dagger3}\hat{\psi}^{3} & = & \left[\left(3\Phi^{3}\Phi^{*}+9n_{c}\tilde{m}+9\Phi^{2}\tilde{n}+9\tilde{n}\tilde{m}\right)\tilde{\psi}^{\dagger}\tilde{\psi}^{\dagger}+\textrm{h.c.}\right]+9\left(n_{c}^{2}+\Phi^{2}\tilde{m}^{*}+\Phi^{*2}\tilde{m}+4n_{c}\tilde{n}+2\tilde{n}^{2}+\tilde{m}\tilde{m}^{*}\right)\tilde{\psi}^{\dagger}\tilde{\psi}\nonumber \\
 &  & +n_{c}^{3}-\left[\left(9\Phi^{2}\tilde{n}\tilde{m}^{*}+\textrm{h.c.}\right)+9n_{c}\left(2\tilde{n}^{2}+\tilde{m}\tilde{m}^{*}\right)+12\tilde{n}^{3}+18\tilde{n}\tilde{m}\tilde{m}^{*}\right].
\end{eqnarray}
By collecting all the \emph{bilinear} terms in the field operators,
we may rewrite the the Hamiltonian density within the Hartree-Fock-Bogoliubov
approximation as,
\begin{equation}
\mathcal{H}_{\textrm{HFB}}=\frac{1}{2}\left[\begin{array}{cc}
\tilde{\psi}^{\dagger}, & \tilde{\psi}\end{array}\right]\left[\begin{array}{cc}
\mathcal{\hat{K}}_{\textrm{HFB}}-\mu & \mathcal{\hat{M}}_{\textrm{HFB}}\\
\mathcal{\hat{M}}_{\textrm{HFB}} & \hat{\mathcal{K}}_{\textrm{HFB}}-\mu
\end{array}\right]\left[\begin{array}{c}
\tilde{\psi}\\
\tilde{\psi}^{\dagger}
\end{array}\right]+\Omega_{\textrm{HFB}}^{(0)}\left(\mathbf{r}\right),\label{eq:HamiHFB}
\end{equation}
where we have defined the operators 
\begin{eqnarray}
\mathcal{\hat{K}}_{\textrm{HFB}} & \equiv & -\frac{\hbar^{2}\nabla^{2}}{2M}+V_{T}+2g\left(n_{c}+\tilde{n}\right)+\frac{3}{2}G\left(n_{c}^{2}+{\color{red}\Phi^{2}\tilde{m}^{*}}+{\color{teal}{\color{red}\Phi^{*2}\tilde{m}}}+4n_{c}\tilde{n}+2\tilde{n}^{2}+{\color{red}\tilde{m}\tilde{m}^{*}}\right),\\
\mathcal{\hat{M}}_{\textrm{HFB}} & \equiv & g\left(\Phi^{2}+{\color{red}\tilde{m}}\right)+G\left(n_{c}\Phi^{2}+{\color{red}3n_{c}\tilde{m}}+3\Phi^{2}\tilde{n}+{\color{red}3n\tilde{m}}\right),
\end{eqnarray}
and 
\begin{eqnarray}
\Omega_{\textrm{HFB}}^{(0)}\left(\mathbf{r}\right) & \equiv & \Phi^{*}\left[-\frac{\hbar^{2}\nabla^{2}}{2M}+V_{T}-\mu\right]\Phi+\frac{g}{2}\left(n_{c}^{2}-2\tilde{n}^{2}-{\color{red}\tilde{m}\tilde{m}^{*}}\right)\nonumber \\
 &  & +\frac{G}{6}\left[n_{c}^{3}-\left({\color{red}9\Phi^{2}\tilde{n}\tilde{m}^{*}+\textrm{h.c.}}\right)-9n_{c}\left(2\tilde{n}^{2}+{\color{red}\tilde{m}\tilde{m}^{*}}\right)-12\tilde{n}^{3}-{\color{red}18\tilde{n}\tilde{m}\tilde{m}^{*}}\right]
\end{eqnarray}
is the density of the \emph{mean-field} thermodynamic potential at
zero temperature. Let us now take the Popov approximation and set
$\tilde{m}=0$ and $\tilde{m}^{*}=0$ in $\mathcal{\hat{K}}_{\textrm{HFB}}$,
$\mathcal{\hat{M}}_{\textrm{HFB}}$ and $\Omega_{\textrm{HFB}}^{(0)}(\mathbf{r})$.
By further assuming a \emph{real} ground-state condensate wave-function
$\Phi(\mathbf{r})$, it is easy to see that $\mathcal{\hat{M}}_{\textrm{HFB}}$
becomes 
\begin{equation}
\hat{\mathcal{M}}=gn_{c}+G\left(n_{c}^{2}+3n_{c}\tilde{n}\right),
\end{equation}
and $\mathcal{\hat{K}}_{\textrm{HFB}}$ can be rewrite as,
\begin{equation}
\hat{\mathcal{K}}=\mathcal{\hat{L}}+\mathcal{\hat{M}},
\end{equation}
where the operator $\mathcal{\hat{L}}$ is defined in Eq. (\ref{eq:GPE-1}).
Also, $\Omega_{\textrm{HFB}}^{(0)}(\mathbf{r})$ takes the form,
\begin{equation}
\Omega^{(0)}\left(\mathbf{r}\right)=\Phi^{*}\left[-\frac{\hbar^{2}\nabla^{2}}{2M}+V_{T}-\mu\right]\Phi+g\left(\frac{n_{c}^{2}}{2}-\tilde{n}^{2}\right)+G\left(\frac{n_{c}^{3}}{6}-3n_{c}\tilde{n}^{2}-2\tilde{n}^{3}\right).
\end{equation}
From the Hamiltonian density for the field operators of non-condensate
atoms, Eq. (\ref{eq:HamiHFB}), we directly write down the coupled
HFB-Popov equations for the $\eta$-th quasi-particle wave-functions
$u_{\eta}$ and $v_{\eta}$ with energy $E_{\eta}>0$,
\begin{equation}
\left[\begin{array}{cc}
\hat{\mathcal{L}}-\mu+\mathcal{\hat{M}} & \mathcal{\hat{M}}\\
\mathcal{\hat{M}} & \hat{\mathcal{L}}-\mu+\mathcal{\hat{M}}
\end{array}\right]\left[\begin{array}{c}
u_{\eta}\left(\mathbf{r}\right)\\
v_{\eta}\left(\mathbf{r}\right)
\end{array}\right]=E_{\eta}\left[\begin{array}{c}
+u_{\eta}\left(\mathbf{r}\right)\\
-v_{\eta}\left(\mathbf{r}\right)
\end{array}\right].\label{eq:HFBP-1}
\end{equation}
\end{widetext}The total thermodynamic potential at finite temperature
$T$ is given by, 
\begin{equation}
\Omega={\color{blue}\Omega_{\textrm{LHY}}^{(0)}}+k_{B}T\sum_{\eta}\ln\left[1-e^{-\beta E_{\eta}}\right]+\int d\mathbf{r}\Omega^{(0)}\left(\mathbf{r}\right),\label{eq:Omega}
\end{equation}
where $\beta=1/(k_{B}T)$ is the inverse temperature and
\begin{equation}
\Omega_{\textrm{LHY}}^{(0)}\equiv-\sum_{\eta}\int d\mathbf{r}E_{\eta}\left|v_{\eta}\mathbf{\left(r\right)}\right|^{2}
\end{equation}
is the contribution of quantum fluctuations to the thermodynamic potential,
the so-called Lee-Huang-Yang (LHY) energy term \cite{Dalfovo1999}.
Once the quasi-particle wave-functions are obtained, the noncondensate
density can be calculated according to,
\begin{equation}
\tilde{n}\left(\mathbf{r}\right)={\color{blue}\tilde{n}_{\textrm{qd}}\left(\mathbf{r}\right)}+\sum_{\eta}\frac{\left|u_{\eta}\left(\mathbf{r}\right)\right|^{2}+\left|v_{\eta}\mathbf{\left(r\right)}\right|^{2}}{e^{\beta E_{\eta}}-1},\label{eq:nt}
\end{equation}
where $\tilde{n}_{\textrm{qd}}(\mathbf{r})=\sum_{\eta}\left|v_{\eta}\mathbf{(r})\right|^{2}$
is the depletion to the condensate arising from quantum fluctuations
and is the only contribution to density at zero temperature.

Eq. (\ref{eq:GPE-1}) and Eq. (\ref{eq:HFBP-1}), together with Eq.
(\ref{eq:nt}) and $n_{c}=\left|\Phi(\mathbf{r})\right|^{2}$, form
a closed set of HFB-Popov equations that should be solved \emph{self-consistently}
\cite{Hutchinson1997}. The chemical potential $\mu$ should be adjusted
to satisfy the number equation $N=\int d\mathbf{r}[n_{c}(\mathbf{r})+\tilde{n}(\mathbf{r})]\equiv N_{c}+N_{\textrm{th}}$.

The HFB-Popov theory has been extensively used to describe weakly
interacting Bose gases at finite temperature \cite{Hutchinson1997}.
The advantages and shortages of such a theory are now well understood.
In particular, it is known that the theory does not provide accurate
descriptions at both zero temperature and at temperatures sufficiently
close to the superfluid transition \cite{Shi1998}. At zero temperature,
this is because the anomalous correlation $\tilde{m}(\mathbf{r})$
neglected in the Popov approximation becomes \emph{comparable} to
the normal correlation, i.e., $\tilde{n}_{\textrm{qd}}(\mathbf{r})$,
if we critically examine the role of zero-temperature quantum fluctuations.
For the consistency of the theory, thus it seems necessary to discard
the quantum depletion $\tilde{n}_{\textrm{qd}}\left(\mathbf{r}\right)$
and the LHY energy term $\Omega_{\textrm{LHY}}^{(0)}$. This treatment
is well justified for a weakly interacting Bose gas with a repulsive
two-body interaction alone, where the quantum depletion (or the LHY
energy) contributes only a few percent to the total density (or the
total energy) \cite{Hutchinson1997}. For a three-body interacting
quantum droplet considered in this work, as we shall discuss below,
we find that the quantum depletion is typically at about 10\%, much
smaller than that of a superfluid helium droplet, where 90\% of atoms
are out of the condensate at $T=0$. In the vicinity of the superfluid
transition $T_{c}$, it is also known that the Popov approximation
predicts a \emph{spurious} first-order superfluid transition for a
\emph{homogeneous} Bose gas, as characterized by a very small jump
(i.e., about a few percent) in the condensate density \cite{Shi1998}.
This spurious feature is not important and does not show up when the
system is confined in a harmonic trap.

\section{Numerical\textit{ }calculations }

At zero temperature, where the noncondensate density $\tilde{n}=0$
as we neglect the quantum depletion, the generalized GP equation Eq.
(\ref{eq:GPE-1}) in free space (i.e., $V_{T}=0$) takes the form,
\begin{equation}
\left[-\frac{\hbar^{2}\nabla^{2}}{2M}+g\left|\Phi\right|^{2}+\frac{G}{2}\left|\Phi\right|^{4}\right]\Phi=\mu\Phi.\label{eq:GPET0}
\end{equation}
As $g<0$ and $G>0$, this GP equation admits a \emph{self-bound}
droplet as the ground state for sufficiently large number of atoms
$N=\int d\mathbf{r}\left|\Phi\right|^{2}\gg1$. To see this, let us
check the case of an infinitely large number of atoms, where we can
safely neglect the surface effect and remove the first kinetic term
(i.e., the $\nabla^{2}$ term). We find that the \emph{bulk} chemical
potential ($n=\left|\Phi\right|^{2}$),
\begin{equation}
\mu\left(n\right)=gn+\frac{G}{2}n^{2},
\end{equation}
and consequently the energy per particle,
\begin{equation}
\frac{\epsilon\left(n\right)}{n}=\frac{g}{2}n+\frac{G}{6}n^{2}.\label{eq:DropletEnergyPerParticle}
\end{equation}
It is clear that the energy per particle acquires a \emph{minimum}
at the equilibrium density,
\begin{equation}
n_{0}=\frac{3}{2}\frac{\left(-g\right)}{G}=6\pi\frac{\left|a\right|}{D},
\end{equation}
at which the pressure $P$ vanishes due to the thermodynamic relation,
\begin{equation}
P=\left[n^{2}\frac{\partial\left(\epsilon/n\right)}{\partial n}\right]_{n=n_{0}}=0.
\end{equation}
The system is therefore self-bound into a droplet state at \emph{zero}
pressure, in equilibrium with the surrounding vacuum. In the absence
of the external harmonic trap, the center density would be fixed to
$n_{0}$, if we neglect the boundary (surface) effect. When we add
the particles to the droplet, the droplet expands and increases its
radius, while keeping its bulk density unchanged.

\subsection{Re-scaled units}

In numerical calculations, it is convenient to introduce the units
for length, density and also energy, and to make the equations dimensionless.
For the density unit, the equilibrium density $n_{0}$ is a natural
choice. To fix the length unit $\xi$, let us \emph{require} that
the three-body term in the energy per particle assumes the following
simple form,
\begin{equation}
\frac{G}{6}n^{2}\rightarrow\frac{\bar{G}}{6}\bar{n}^{2}=\bar{n}^{2},
\end{equation}
after we use the re-scaled units (as indicated by the bar above the
variables). This means that the dimensionless three-body interaction
strength is,
\begin{equation}
\bar{G}\equiv\frac{Gn_{0}^{2}}{\hbar^{2}/\left(M\xi^{2}\right)}=+6,
\end{equation}
where $\hbar^{2}/(M\xi^{2})$ is the energy unit related to the length
unit $\xi$. By substituting $G=\hbar^{2}D/M$ and the equilibrium
density $n_{0}$, we find that,
\begin{equation}
\xi=\sqrt{\frac{D}{6\pi^{2}a^{2}}}.
\end{equation}
It is straightforward to check that the two-body interaction strength
then becomes, 
\begin{equation}
\bar{g}\equiv\frac{gn_{0}}{\hbar^{2}/\left(M\xi^{2}\right)}=-4.
\end{equation}
From now on, we will use the \emph{re-scaled} units for length $\mathbf{\bar{r}}\equiv\mathbf{r}/\xi$,
density $\bar{n}\equiv n/n_{0}$, energy $\bar{E}_{\eta}\equiv M\xi^{2}E_{\eta}/\hbar^{2}$
and trapping frequency $\bar{\omega}=M\xi^{2}\omega/\hbar$. The re-scaled
condensate wave-function is $\bar{\Phi}=\Phi/\sqrt{n_{0}}$. The reduced
number of particles is given by $\bar{N}=N/(n_{0}\xi^{3})$ with 
\begin{equation}
n_{0}\xi^{3}=\frac{\sqrt{D/6}}{\pi^{2}a^{2}}.
\end{equation}
and the number equation becomes, 
\begin{equation}
\bar{N}=\int d\mathbf{\bar{r}}\frac{n_{c}\left(\mathbf{\bar{r}}\right)+\tilde{n}\left(\mathbf{\bar{r}}\right)}{n_{0}}\equiv\bar{N}_{c}+\bar{N}_{\textrm{th}}.
\end{equation}
\textcolor{red}{Hereafter, for convenience we shall }\textcolor{red}{\emph{remove}}\textcolor{red}{{}
the bar for all the re-scaled units}. 

In the dimensionless form ($g=-4$ and $G=+6$, and effectively $\hbar=M=1$),
the zero-temperature GP equation Eq. (\ref{eq:GPET0}) then becomes
very simple, 
\begin{equation}
\left[-\frac{\nabla^{2}}{2}+\frac{\omega^{2}r^{2}}{2}-4\Phi^{2}+3\Phi^{4}\right]\Phi=\mu\Phi,\label{eq:GPET0s}
\end{equation}
where we have re-inserted the harmonic trap term. We see that this
equation depends on two \emph{controlling} variables, the dimensionless
trapping frequency $\omega$ and the reduced number of particle $N$,
both of which involves the microscopic parameters of the model Hamiltonian
such as the two-body scattering length $a$ and the three-body scattering
hypervolume $D$. Eq. (\ref{eq:GPET0s}) has been discussed in detail
in the previous work \cite{Hu2020c}. It turns out that for a large
reduced number of particles, we can neglect the finite-size effect
and the properties of the system actually depend on a single parameter
$N^{1/3}\omega$. In our calculations, we typically take the reduced
number of particles $N=1000$.

\subsection{Technical difficulties}

The dimensionless HPB-Popov equations with the operators $\hat{\mathcal{L}}$
and $\hat{\mathcal{M}}$, obtained by setting $\hbar=M=1$, $g=-4$
and $G=+6$, can be solved by the routines outlined in the previous
work \cite{Hutchinson1997,Hu2020c,Wang2020b}. The following three
challenges in the numerical calculations are worth noting.

First, for a droplet state, there are numerous quasiparticle energy
levels accumulated just above the particle emission threshold $\left|\mu\right|$
\cite{Hu2020c}. The energy level separation is set by the harmonic
trapping frequency $\omega$. To solve this difficulty, we consider
a dimensionless trapping frequency $\omega\geq0.03$ and typically
use several-hundred expansion basis functions in solving the quasi-particle
wave-functions $u_{\eta}$ and $v_{\eta}$ for a given angular momentum
$l$ (which is a good quantum number for our spherical harmonic traps).
Further improvement on the numerical accuracy could be achieved by
considering the 5-th order $B$-spline basis \cite{Wang2020b}. The
$B$-spline basis would allow us to use an uneven grid, which might
better represent the solution wave-function. It also allows a higher
order approximation of the derivative operator that appears in the
kinetic energy term.

On the other hand, at finite temperature $T\neq0$ we must \emph{iterate}
the solutions of the GP equation Eq. (\ref{eq:GPE-1}) for the condensate
wave-function and the HFB-Popov equations Eq. (\ref{eq:HFBP-1}) for
the quasi-particle wave-functions, in order to gradually improve the
thermal density $\tilde{n}(\mathbf{r})$ for convergence. This iterative
procedure turns out to be very slow with increasing temperature. In
particular, sufficiently close to the superfluid transition, the small
number of condensed particles $N_{c}\sim O(1)$ implies that the chemical
potential $\mu$ appearing in the GP equation (i.e., the lowest eigenvalue
of the operator $\mathcal{\hat{L}}$ in the zero momentum $l=0$ sector)
can no longer be treated as the chemical potential of the \emph{whole}
system. To solve this problem, we introduce a new chemical potential
$\mu_{t}$ of the whole system, by requiring that,
\begin{equation}
\frac{1}{e^{\left(\mu-\mu_{t}\right)/k_{B}T}-1}=N_{c}.
\end{equation}
The difference between the two chemical potentials 
\begin{equation}
\mu-\mu_{t}=k_{B}T\ln\left(1+\frac{1}{N_{c}}\right)\sim\frac{k_{B}T}{N_{c}}
\end{equation}
is negligibly small away from the superfluid transition when $N_{c}\gg1$.
However, in the vicinity of the superfluid transition, in the Bose-Einstein
distribution function $f_{B}(E_{\eta}$) we have to measure the energy
$E_{\eta}$ of the Bogoliubov quasiparticles with respect to $\mu_{t}$,
instead of $\mu$. This leads to a modified expression for the thermal
density (in the case of excluding the quantum depletion), 
\begin{eqnarray}
\tilde{n}\left(\mathbf{r}\right) & = & \sum_{\eta}\frac{\left|u_{\eta}\left(\mathbf{r}\right)\right|^{2}+\left|v_{\eta}\mathbf{\left(r\right)}\right|^{2}}{e^{\beta\left(E_{\eta}+\mu-\mu_{t}\right)}-1}\nonumber \\
 & = & \sum_{\eta}\frac{\left|u_{\eta}\left(\mathbf{r}\right)\right|^{2}+\left|v_{\eta}\mathbf{\left(r\right)}\right|^{2}}{\left(1+1/N_{c}\right)e^{\beta E_{\eta}}-1}.
\end{eqnarray}
The thermal contribution to the thermodynamic potential in Eq. (\ref{eq:Omega})
is modified in a similar way (i.e., by replacing $E_{\eta}$ with
$E_{\eta}+\mu-\mu_{t}$).

Finally, a large temperature leads to the significant population of
the \emph{high-energy} quasi-particle energy levels, which would require
a large number of the expansion basis functions in solving the HFB-Popov
equations and therefore considerably slow down our numerical calculations.
To avoid this problem, we consider the use of the local-density approximation
(LDA) for the high-lying quasi-particle wave-functions \cite{Liu2007}.
For this purpose, we introduce a high-energy cut-off $E_{c}$. For
the quasi-particle energy $E_{\eta}>E_{c}$, we treat the quasi-particle
wave-functions locally (at the position $r$) as plane-waves with
amplitudes $u_{\mathbf{k}}(r)$ and $v_{\mathbf{k}}(r)$ at the momentum
$\mathbf{k}$ \cite{Liu2007}. The HFB-Popov equations for the quasi-particles
then become,
\begin{equation}
\left[\begin{array}{cc}
\frac{k^{2}}{2}+\mathcal{V}_{\textrm{eff}}(r) & \mathcal{M}(r)\\
\mathcal{M}(r) & \frac{k^{2}}{2}+\mathcal{V}_{\textrm{eff}}(r)
\end{array}\right]\left[\begin{array}{c}
u_{\mathbf{k}}\\
v_{\mathbf{k}}
\end{array}\right]=E_{\mathbf{k}}\left[\begin{array}{c}
+u_{\mathbf{k}}\\
-v_{\mathbf{k}}
\end{array}\right],
\end{equation}
where 
\begin{eqnarray}
\mathcal{V}_{\textrm{eff}} & = & \frac{\omega^{2}r^{2}}{2}-\mu-8\left(n_{c}+\tilde{n}\right)+9n_{c}^{2}+36n_{c}\tilde{n}+6\tilde{n}^{2},\\
\mathcal{M} & = & -4n_{c}+6n_{c}^{2}+18n_{c}\tilde{n},
\end{eqnarray}
from which we obtain,
\begin{eqnarray}
u_{\mathbf{k}}^{2}(r) & = & \frac{1}{2}\left[\frac{k^{2}/2+\mathcal{V}_{\textrm{eff}}\left(r\right)}{E_{\mathbf{k}}\left(r\right)}+1\right],\\
v_{\mathbf{k}}^{2}(r) & = & \frac{1}{2}\left[\frac{k^{2}/2+\mathcal{V}_{\textrm{eff}}\left(r\right)}{E_{\mathbf{k}}\left(r\right)}-1\right],
\end{eqnarray}
and 
\begin{equation}
E_{\mathbf{k}}(r)=\sqrt{\left[k^{2}/2+\mathcal{V}_{\textrm{eff}}(r)\right]^{2}-\mathcal{M}^{2}(r)}.
\end{equation}
At the position $r$, therefore the LDA contribution from the \emph{continuous}
high-lying energy levels to the thermal density (i.e., $E_{\mathbf{k}}(r)>E_{c}$)
is, 
\begin{equation}
\tilde{n}_{H}\left(\mathbf{r}\right)=\int\frac{d\mathbf{k}}{\left(2\pi\right)^{3}}\frac{\left[\frac{k^{2}}{2}+\mathcal{V}_{\textrm{eff}}(r)\right]/E_{\mathbf{k}}(r)}{\left(1+1/N_{c}\right)e^{\beta E_{\mathbf{k}}(r)}-1}.
\end{equation}
Together with the contribution from the \emph{discrete} low-lying
energy levels, 
\begin{equation}
\tilde{n}_{L}\left(\mathbf{r}\right)=\sum_{E_{\eta}<E_{c}}\frac{\left|u_{\eta}\left(\mathbf{r}\right)\right|^{2}+\left|v_{\eta}\mathbf{\left(r\right)}\right|^{2}}{\left(1+1/N_{c}\right)e^{\beta E_{\eta}}-1},
\end{equation}
we obtain the total thermal density,
\begin{equation}
\tilde{n}(\mathbf{r})=\tilde{n}_{L}\left(\mathbf{r}\right)+\tilde{n}_{H}\left(\mathbf{r}\right).
\end{equation}
The LDA treatment for the high-lying quasi-particle energy levels
turns out to be very efficient. Our numerical results are essentially
independent on the cut-off energy $E_{c}$, provided that it is reasonably
large.

\section{Quantum depletion}

\begin{figure}[t]
\centering{}\includegraphics[width=0.45\textwidth]{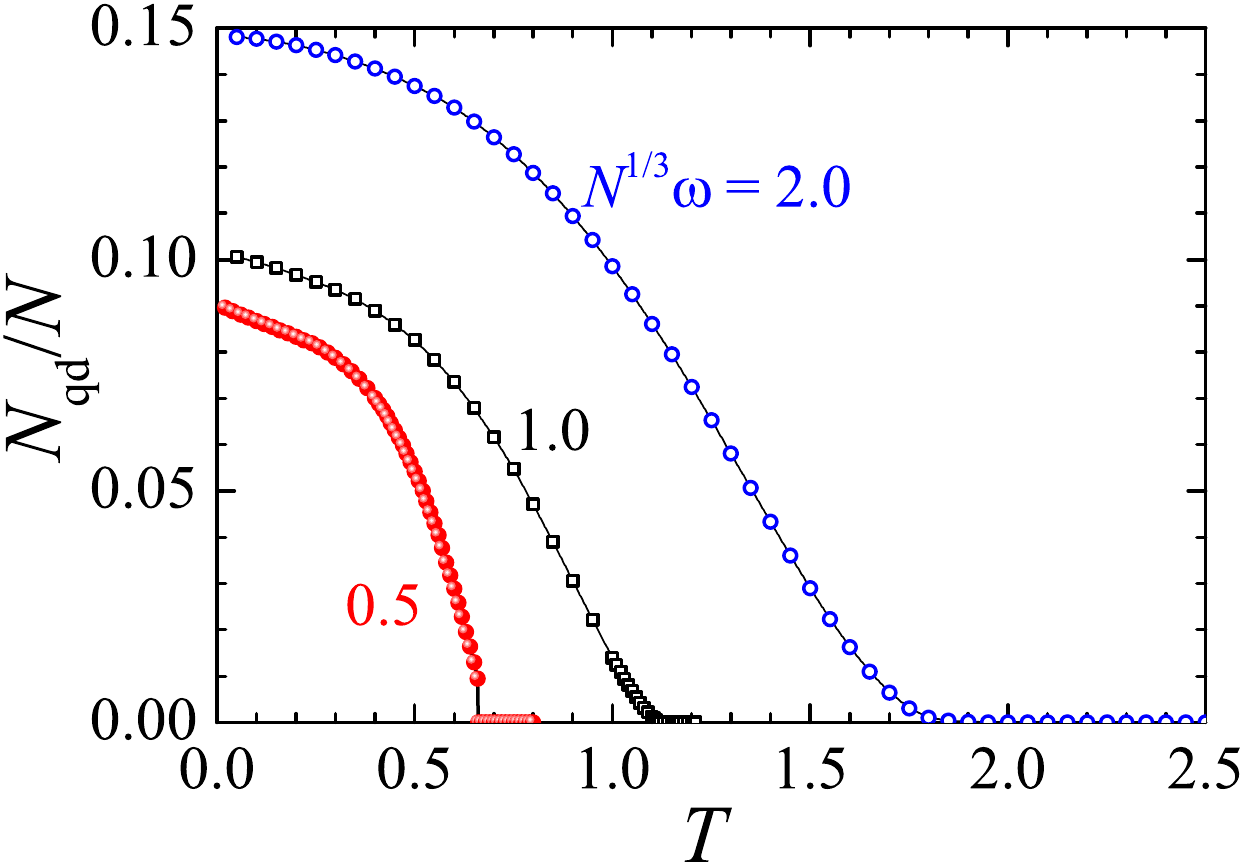} \caption{\label{figS1_Nqd} Quantum depletion $N_{\textrm{qd}}/N$ as a function
of temperature at three effective trapping frequencies $N^{1/3}\omega=0.5$,
$1.0$ and $2.0$. The depletion is typically about $10\%$. It vanishes
when we increase the temperature towards the superfluid transition.}
\end{figure}

We have self-consistently (and iteratively) solved the coupled HFB-Popov
equations, \emph{without} the inclusion of the quantum depletion $\tilde{n}_{\textrm{qd}}(\mathbf{r})$
in the thermal density $\tilde{n}(\mathbf{r})$. We have then calculated
the number of quantum depleted atoms $N_{\textrm{qd}}=\int d\mathbf{r}\sum_{\eta}\left|v_{\eta}\mathbf{(r})\right|^{2}$
for a self-consistent check. The results are shown in Fig. \ref{figS1_Nqd},
where we consider the temperature dependence of the ratio $N_{\textrm{qd}}/N$
at three effective trapping frequencies. We have also tried the inclusion
of $\tilde{n}_{\textrm{qd}}(\mathbf{r})$ in the thermal density $\tilde{n}(\mathbf{r})$
in our numerical iterations and have found qualitatively similar predictions.

It is readily seen that the quantum depletion is typically at around
$10\%$ at zero temperature. Actually, we would anticipate that the
zero-temperature quantum depletion should be a \emph{universal} constant
in the absence of the external harmonic trap (i.e., about $8\%$),
where the bulk density of the droplet state is precisely the equilibrium
density $n_{0}$. The increase of the zero-temperature quantum depletion
with increasing effective trapping frequency should be contributed
to the enhanced central density due to the external trapping potential.

At nonzero temperature, the quantum depletion decreases as anticipated.
It becomes negligibly small close to the superfluid transition. As
in this work we are mostly interested in the region near the superfluid
transition, the consistent exclusion of the quantum depletion in our
HFB-Popov calculations therefore seems well justified. 

\section{First-order BEC transition temperature}

The first-order BEC transition temperature at \emph{small} effective
trapping frequency $N^{1/3}\omega\rightarrow0$ might be analytically
derived. To see this, let us consider the free energy $F_{S}(T)$
of the superfluid quantum droplet state and the free energy $F_{N}(T)$
of a normal Bose gas.

\subsection{The normal-state free energy $F_{N}(T)$}

For the normal phase, since the trapping frequency is very small,
the atomic cloud is dilute and can be treated as a non-interacting
Bose gas. The free energy can then be obtained in the semi-classical
approximation \cite{Dalfovo1999},
\begin{equation}
F_{N}=\iint\frac{d\mathbf{k}d\mathbf{r}}{\left(2\pi\right)^{3}}\ln\left[1-e^{-\beta\left(\frac{\hbar^{2}k^{2}}{2m}+V_{T}-\mu\right)}\right]+\mu N,
\end{equation}
where the chemical potential $\mu$ satisfies, 
\begin{equation}
N=\iint\frac{d\mathbf{k}d\mathbf{r}}{\left(2\pi\right)^{3}}\frac{1}{e^{\beta\left(\frac{\hbar^{2}k^{2}}{2m}+V_{T}-\mu\right)}-1}.
\end{equation}
These Bose-type integrals can be worked out explicitly, with the help
of the polylogarithm function $\textrm{Li}_{\nu}(z)$, i.e., 
\begin{equation}
\int_{0}^{\infty}d\epsilon\frac{\epsilon^{\nu-1}}{e^{\beta\left(\epsilon-\mu\right)-1}}=\left(k_{B}T\right)^{\nu}\Gamma\left(\nu\right)\textrm{Li}_{\nu}\left(z\right),
\end{equation}
where the fugacity $z\equiv e^{\beta\mu}$ and $\Gamma(\nu)$ is the
Gamma function. We find that \cite{Dalfovo1999},
\begin{eqnarray}
\frac{F_{N}}{N} & = & k_{B}T\frac{\textrm{Li}_{4}\left(z\right)}{\textrm{Li}_{3}\left(z\right)}+\mu,\\
N & = & \left(\frac{k_{B}T}{\hbar\omega}\right)^{3}\textrm{Li}_{3}\left(z\right).
\end{eqnarray}
In the normal phase, the fugacity $z$ turns out to be very small.
Indeed, below the condensation transition the chemical potential $\mu$
must turn into the bulk chemical potential of the droplet state, i.e.,
$\mu_{0}=gn_{0}+Gn_{0}^{2}/2<0$, which serves as the energy unit
after we re-scale and make the equations dimensionless. For example,
by taking $g=-4$, $G=+6$ and $n_{0}=1$, we obtain $\mu_{0}=-1$.
Therefore, it is reasonable to Taylor expand the expressions of the
free energy and the number of particles, in terms of the small fugacity
$z\sim e^{\mu_{0}/(k_{B}T)}\ll1$ at low temperature $k_{B}T\ll\left|\mu_{0}\right|$.
In the \emph{re-scaled} units, after we take $k_{B}=1$ and $\hbar=1$,
we then obtain that,
\begin{eqnarray}
\frac{F_{N}}{N\left|\mu_{0}\right|} & \simeq & T\left(1-\frac{z}{16}\right)+T\ln z,\\
N\omega^{3} & \simeq & T^{3}z\left(1+\frac{z}{8}\right).
\end{eqnarray}
By re-expressing the fugacity $z$ in terms of the temperature $T$,
to the leading order we finally arrive at, 
\begin{equation}
\frac{F_{N}}{N\left|\mu_{0}\right|}\simeq3T\ln\frac{N^{1/3}\omega}{T}.
\end{equation}

\subsection{The droplet-state free energy $F_{S}(T)$}

At low temperature, the free energy $F_{S}(T)$ is essentially the
zero-temperature total energy $E(T=0)$ of the droplet, if we neglect
the small surface effect for a large droplet (with $N\gg1$). By using
Eq. (\ref{eq:DropletEnergyPerParticle}), in the re-scaled units we
find that,
\begin{equation}
\frac{F_{S}}{N\left|\mu_{0}\right|}\simeq-1.
\end{equation}
This expression is easy to understand. At zero temperature we anticipate
that all the atoms are bounded into the droplet with the binding energy
$\left|\mu_{0}\right|$.

\subsection{First-order superfluid transition temperature}

Using the condition $F_{N}(T)=F_{S}(T)$ at the transition temperature
$T_{c}$, we obtain
\begin{equation}
3T_{c}\ln\frac{N^{1/3}\omega}{T_{c}}=-1,
\end{equation}
or
\begin{equation}
N^{1/3}\omega=T_{c}\exp\left(-\frac{1}{3T_{c}}\right).
\end{equation}
By restoring the full units, we find that,
\begin{equation}
N^{1/3}\hbar\omega=k_{B}T_{c}\exp\left(\frac{\mu_{0}}{3k_{B}T_{c}}\right),
\end{equation}
where the bulk chemical potential takes the form,
\begin{equation}
\mu_{0}=-\frac{3g^{2}}{8G}.
\end{equation}
It is clear that the critical temperature eventually goes to zero,
$T_{c}\rightarrow0$, when we gradually remove the external harmonic
trapping potential, $\omega\rightarrow0$.

\begin{figure}[t]
\centering{}\includegraphics[width=0.45\textwidth]{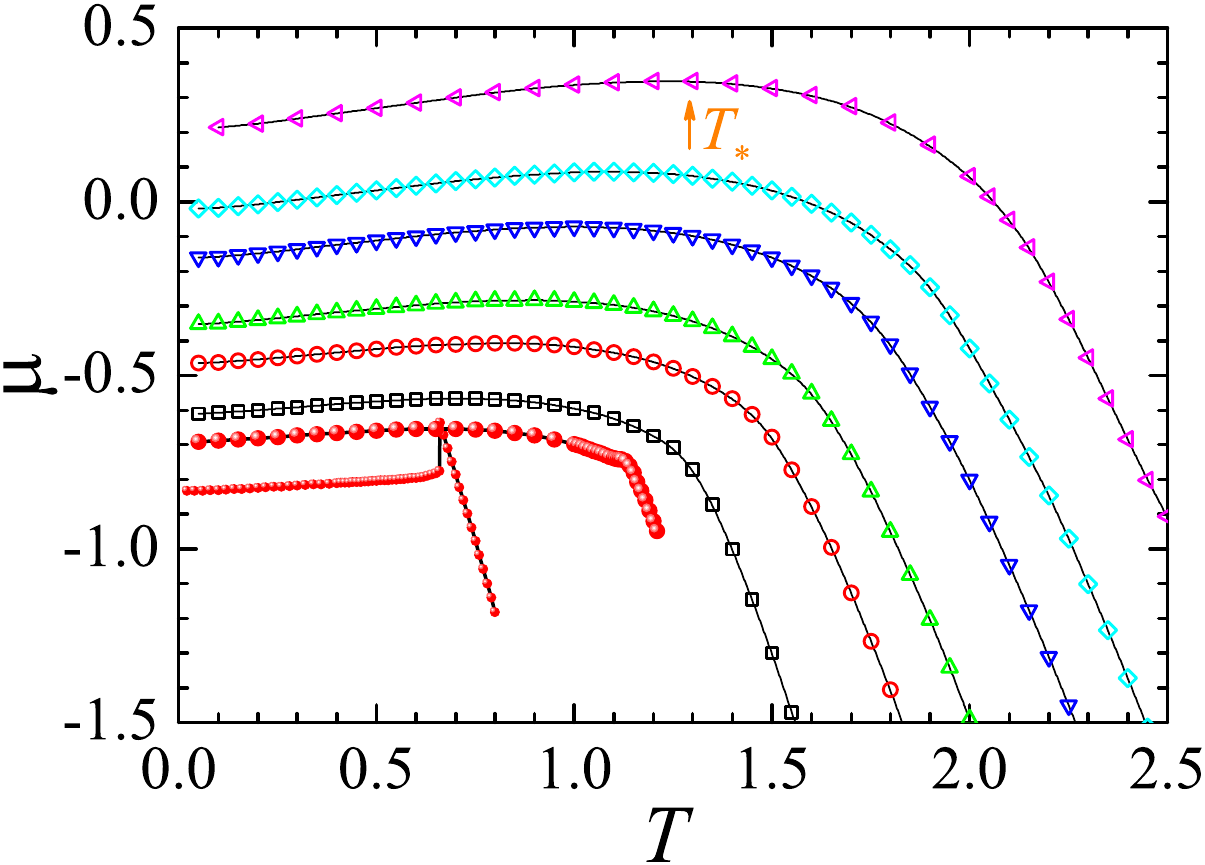} \caption{\label{figS2_mu} Temperature dependence of the chemical potential
at different effective trapping frequency $N^{1/3}\omega$. From bottom
to top, the value of $N^{1/3}\omega$ is $0.5$, $1.0$, $1.2$, $1.5$,
$1.7$, $2.0$, $2.2$, and $2.5$.}
\end{figure}

\section{The characteristic temperature $T_{*}$ from the chemical potential}

At large effective trapping frequency $N^{1/3}\omega>1.0$, with decreasing
temperature the cloud of three-body interacting bosons experiences
a conventional smooth second-order phase transition from a normal
gas to a superfluid gas at $T_{c}$. If we further decrease the temperature
below $T_{c}$, the system may become a superfluid liquid (i.e., a
quantum droplet) at a characteristic temperature $T_{*}$, where the
chemical potential acquires a \emph{maximum} in its temperature dependence. 

This is not always true, when we decrease the effective trapping frequency
to the regime $N^{1/3}\omega<1.0$, as reported in Fig. \ref{figS2_mu}.
In that regime, the temperature dependence of the chemical potential
becomes rather flattened. Moreover, at even smaller effective trapping
frequency (i.e., the lowest curve with $N^{1/3}\omega=0.5$), the
chemical potential monotonically increases with increasing temperature,
before the sudden (first-order) jump to the chemical potential of
a normal phase. We can not observe that the characteristic temperature
$T_{*}$ smoothly merges with the transition temperature $T_{c}$
at the tri-critical point $(N^{1/3}\omega,T_{c})_{\textrm{tri}}\simeq(1.0,1.13)$.

\begin{figure*}
\begin{centering}
\includegraphics[width=0.9\textwidth]{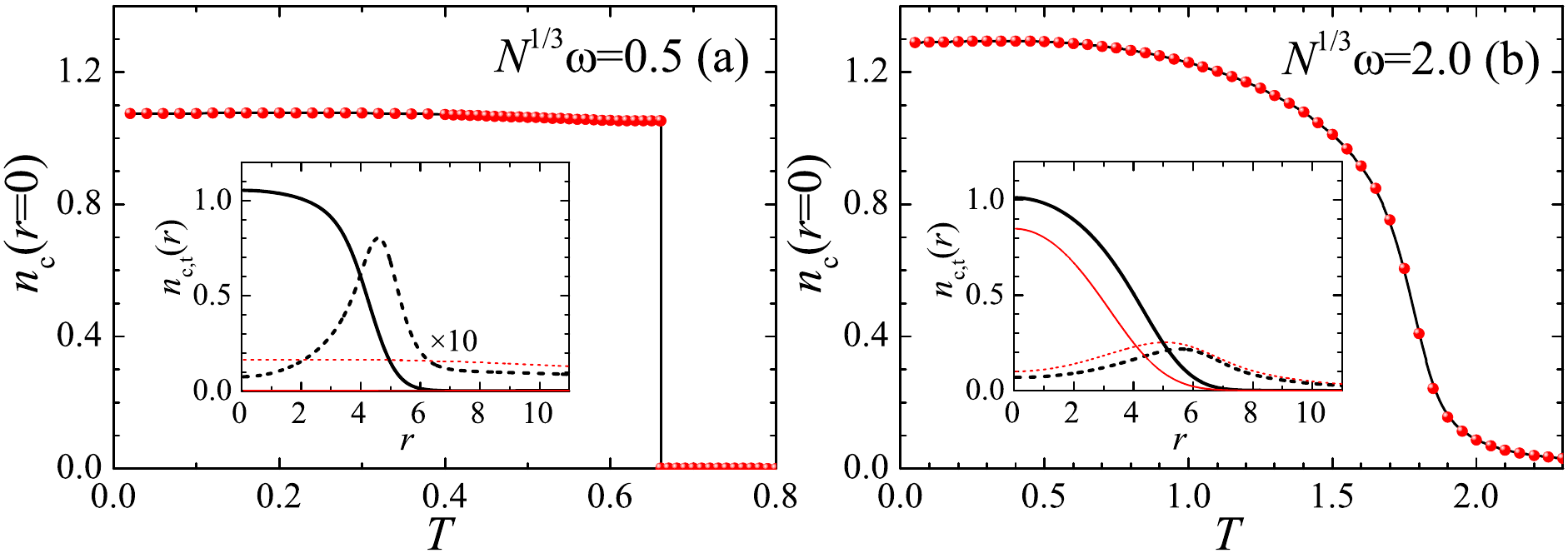}
\par\end{centering}
\caption{\label{figS3_density} Temperature dependence of the central condensate
density at $N^{1/3}\omega=0.5$ (a) and $N^{1/3}\omega=2.0$ (b).
The insets show the condensate and thermal density distributions at
$T=0.95T_{c}$ (black thick curves) and at $T=1.05T_{c}$ (red thin
curves), respectively. The condensate distribution is plotted in solid
line and the thermal distribution is in dashed line.}
\end{figure*}

\section{Central condensate density across the superfluid transition}

The difference in the first-order and second-order BEC transitions
can be mostly easily identified by measuring the central condensate
density at the trap center. In Fig. \ref{figS3_density}, we report
the central condensate density as a function of temperature at two
different effective trapping frequencies $N^{1/3}\omega=0.5$ (a)
and $N^{1/3}\omega=2.0$ (b), at which the system experiences the
first-order transition and second-order transition, respectively.
The two insets show the spatial distributions of the condensate density
and the thermal density before and after the BEC transition, see,
for example, the black thick curves at $T=0.95T_{c}$ and the red
thin curves at $T=1.05T_{c}$.

The first-order BEC transition is clearly revealed in the inset of
Fig. \ref{figS3_density}(a), where the central condensate density
suddenly jumps to zero, while the thermal density (enlarged by ten
times in the inset) becomes nearly flat in the region plotted. In
sharp contrast, at second-order BEC transition the central condensate
density gradually vanishes as the temperature increases. As illustrated
in the inset of Fig. \ref{figS3_density}(b), there is still a considerable
condensate density at $T=1.05T_{c}$, due to the finite number of
particles (i.e., the finite-size effect).

\begin{thebibliography}{10}
\bibitem{muQuantumDroplet}The many-body binding energy of a quantum
droplet is given by $-\mu$. We may anticipate the binding energy
of the droplet becomes smaller with increasing temperature. Hence,
the chemical potential of the droplet increases as temperature increases.
\end{thebibliography}
\end{document}